\documentclass[]{aa}

\usepackage{amsmath}
\usepackage{graphicx}
\usepackage{graphics}
\usepackage{epsfig}
\usepackage{txfonts}
\begin{document}

\title{The High A$_V$ Quasar Survey: A $z=2.027$ metal-rich damped
Lyman-$\alpha$ absorber towards a red quasar at $z=3.21$ 
} \titlerunning{A $z=2.027$ metal-rich
damped Lyman-$\alpha$ absorber
}

\author{
J.~P.~U.~Fynbo\inst{1},
J.-K.~Krogager\inst{1, 2},
K.~E.~Heintz\inst{3,1},
S.~Geier\inst{4,5},
P.~M\o ller\inst{6},
P.~Noterdaeme\inst{2},
L.~Christensen\inst{1},
C.~Ledoux\inst{7},
P. Jakobsson\inst{3}
}
\institute{
Dark Cosmology Centre, Niels Bohr Institute, University of Copenhagen, Juliane
Maries Vej 30, 2100 Copenhagen \O, Denmark
\email{jfynbo@nbi.ku.dk}
\and
Institut d'Astrophysique de Paris, CNRS-UPMC, UMR7095, 98bis bd Arago, F-75014 Paris, France
\and
Centre for Astrophysics and Cosmology, Science Institute, University of Iceland, Dunhagi 5, 107 Reykjav\'ik, Iceland
\and
Instituto de Astrof{\'i}sica de Canarias, V{\'i}a L{\'a}ctea, s/n, 38205, La Laguna, Tenerife, Spain
\and
Gran Telescopio Canarias (GRANTECAN), Cuesta de San Jos{\'e} s/n, E-38712, Bre{\~n}a Baja, La Palma, Spain
\and
European Southern Observatory, Karl-Schwarzschildstrasse 2, D-85748 Garching, Germany
\and
European Southern Observatory, Alonso de C{\'o}rdova 3107, Casilla 19001, Vitacura, Santiago 19, Chile
}
\authorrunning{Fynbo et al.}

\date{Received 2017; accepted, 2017}

\abstract{
It is important to
understand the selection effects behind the quasar samples to fully exploit the potential of quasars as probes of cosmic chemical
evolution and the internal gas dynamics of galaxies; in particular, it is vital to understand
whether the selection criteria exclude foreground galaxies with certain properties,
most importantly a high dust content. Here we present spectroscopic follow-up
from the 10.4 m GTC telescope of a dust-reddened quasar,
eHAQ0111+0641, from the extended High A$_\mathrm{V}$ Quasar (HAQ) survey.  We
find that the $z=3.21$ quasar has a foreground damped Lyman-$\alpha$ absorber
(DLA) at $z=2.027$ along the line of sight. The DLA has very strong metal lines
due to a moderately high metallicity with an inferred lower limit of 25\% of
the solar metallicity, but a very large gas column density along the
line of sight in its host galaxy.  This discovery is further evidence that
there is a dust bias affecting the census of metals, caused by the combined
effect of dust obscuration and reddening, in existing samples of $z>2$ DLAs.
The case of eHAQ0111+0641 illustrates that dust bias is not only caused by dust
obscuration, but also dust reddening.
}

\maketitle

\section{Introduction}     
\label{sec:introduction}

During the last two decades the study of the galaxy population of the first
few Gyr after the Big Bang has undergone a revolution from a state with nearly no data
to a state today with thousands of galaxies identified and studied in emission
over a range of redshifts extending back to $z>10$ \citep{Madau2014,Stark2016}.
Prior to this revolution, galaxies at these early epochs were only studied in
absorption against bright backlights such as quasars \citep{Weymann1981}. 
The DLAs, which are Ly$\alpha$ absorbers with
\ion{H}{i} column densities above $10^{20.3}$ cm$^{-2}$, are of particular
interest as they have all the properties expected for sightlines transversing
the interstellar medium or, at least, circumgalactic material of actual galaxies \citep{Wolfe1986,Wolfe2005}.

Connecting the information inferred from studies of high redshift galaxies via
absorption line studies and direct emission has not been straightforward.
The so-called galaxy counterparts of the absorption selected 
systems, i.e. the shining components of the objects hosting the gas responsible 
for absorption line systems seen in quasar spectra, have long remained elusive
to detect \citep[e.g.][and references therein]{Moller93,Djorgovski1996,Fynbo2010}. 

A consistent picture is now emerging in which absorption, and here we
specifically refer to DLAs, and emission studies of galaxies at $z>2$ seem to
probe the same underlying galaxy population in a way that to first order can be
understood based on simple power-law scaling relations (evolving with redshift)
between mass, luminosity, size, and metallicity
\citep[e.g.][]{Fynbo1999,Haehnelt2000,Schaye2001,Moller2002,Moller2004,ledoux2006,
Fynbo2008,Krogager2012,Moller2013,Neeleman2013,Christensen2014,Krogager2017}.
In this picture the most metal-rich DLAs are formed in large and luminous
galaxies, whereas the typical DLAs with lower metallicities are predominantly
formed in much more numerous, but smaller and fainter dwarf galaxies. A good
illustration of this is in \citet[][their Fig.~10]{Krogager2017}. At least
some hydro simulations of galaxies at these redshifts confirm this picture
\citep{Pontzen2008,Rahmati2014}.

However, an important caveat to keep in mind is the issue of dust bias in the
DLA samples.  Metal-strong and DLAs  that are, hence, likely to be dusty decrease the
detection probability of the background quasars and hence such systems are
under-represented in quasar samples selected in the optical. 
This effect has
been discussed intensively in the literature
\citep[e.g.][]{Ostriker1984,Pei1991,Boisse1998,Pei1999,Vladilo2005,Trenti2006,Pontzen2009,Wang2012}. 

Absorption statistics from radio selected samples of QSOs are free from dust
bias as radio emission is not affected by dust and {\it this} approach has been followed
both in the CORALS survey \citep{Ellison2001,Ellison2005} and the UCSD
survey \citep{Jorgenson2006}. The largest of those studies is the UCSD survey,
which includes the CORALS data, but for which the optical identification is not
complete. In this survey 26 DLAs have been identified, but
\citet{Jorgenson2006} argue that a survey that is four times larger is needed to make a
firm conclusion about the presence of a dust bias.  \citet{Pontzen2009}
argued that dust-bias is likely to be a small effect, but they found that
the cosmic density of metals as measured from DLA surveys could be
underestimated by as much as a factor of 2. Hence, further exploring and
establishing the magnitude of the possible dust bias is important for the issue
of cosmic chemical evolution. 

The effect of dust that has been analysed in the context of DLAs and cosmic
chemical
evolution is that of obscuration of the light from the background quasar
\citep[e.g.][]{Fall1993,Boisse1998,Pei1999,Smette2005,Vladilo2005,Trenti2006,Pontzen2009}.
The other important effect of the dust is reddening, which also decreases the
detection probability of the background quasar.  Tentative evidence for this
effect was found in a study of a metal-rich and dusty DLA in
\citet{Fynbo2011}. In this case the DLA moved the background quasar out to
the edge of the colour distribution of the SDSS quasar locus and the DLA was only
selected for spectroscopy because it fell into the selection window for high-$z$
quasars because of its faint u-band flux.  Several other quasars reddened by a
foreground dusty absorbers have been found in SDSS, but again only because
the strong reddening made the quasars fall into the selection window for very
high redshift quasars or as filler objects for unused fibers
\citep[e.g.][]{Noterdaeme2009,Wang2012,Pan2017}. 

Motivated by the object studied in \citet{Fynbo2011} the High A$_\mathrm{V}$
Quasar Survey \cite[HAQ;][]{Fynbo2013,Krogager2015,Krogager2016b} for
reddened quasars was initiated in 2011 with the objective of measuring the
frequency of sources that are reddened out of the quasar selection criteria
typically adopted in surveys such as SDSS \citep[e.g.][]{Schneider2010}. In the
HAQ survey we selected candidate quasars as point sources in regions of the sky
with coverage at both optical and near-IR wavelengths, such as the overlapping
regions of the Sloan Digital Sky Survey (SDSS) and UKIRT Infrared Deep Sky Survey (UKIDSS) 
\citep{Warren2007,Eisenstein2011}. 
The HAQ survey has revealed a multitude of red quasars, 
but the large majority, more than 90\%, of these
quasars are red for other reasons than dust in foreground DLAs \citep[see
also][]{Krawczyk2015,Zafar2015}.

\citet{Krogager2016a} have presented the first clear case of a very bright quasar
from the HAQ survey reddened by a $z>2$ foreground, metal-rich, and dusty DLA.
In this paper we present the second detection.  eHAQ0111+0641 was selected
from SDSS and UKIDSS imaging as part of the extended HAQ survey
\citep{Krogager2016b}. In SDSS it is listed as an object of type 'Star' based
on its optical photometry.  Its celestial coordinates are (J2000.0) RA =
01:11:34.7, Dec = +06:41:19.2. \citet{Krogager2016b} already noted this
as a system where there is evidence for the 2175-\AA \ dust extinction
feature at a fitted redshift of 2.04 consistent within the errors with a
strong metal-line system at $z=2.027$.

The paper is organized in the following way: in Sect.~\ref{sec:data} we
present our new spectroscopic observations of eHAQ0111+0641. In 
Sect.~\ref{sec:results} we present our results on the details of the DLA,
constraints on its metallicity, and on the extinction. Finally we offer
our conclusions in Sect.~\ref{sec:conc}.

\section{Observations and data reduction}    \label{sec:data}

eHAQ0111+0641 was observed with the OSIRIS instrument at the 
Gran Telescopio Canarias (GTC) as part of
a larger sample of candidate red quasars. We
secured spectroscopy with OSIRIS and a range of grisms
to better constrain the spectral energy distribution, metal
lines, and hydrogen Ly$\alpha$ line. The log of observations is provided
in Table~\ref{tab:log}.

\begin{table}[!htbp]
\centering
\begin{minipage}{0.49\textwidth}
\centering
\caption{Log of observations.}
\begin{tabular}{lrrrccccccccc}
\noalign{\smallskip} \hline \hline \noalign{\smallskip}
Date & Grism & Resolution & Exptime & Airmass \\
     &       &            &  (sec)  &      \\
\hline
11/09/2016 & 1000R & 450 & 2$\times$500  & 1.34--1.37 \\
26/09/2016 & 1000B & 500 & 3$\times$1000 & 1.22--1.33 \\
09/10/2016 & 2500U & 1250 & 4$\times$1800 & 1.11-1.08 \\ 
19/01/2017 & 2500R & 1850 & 3$\times$720 & 1.32--1.43 \\
\noalign{\smallskip} \hline \noalign{\smallskip}
\end{tabular}
\centering
\label{tab:log}
\end{minipage}
\end{table}

The spectroscopic data were reduced using standard procedures in 
IRAF.\footnote{IRAF is distributed by the National Optical Astronomy
Observatory, which is operated by the Association of Universities for
Research in Astronomy (AURA) under a cooperative agreement with the
National Science Foundation.} The spectra were flux calibrated with the 
observations of spectro-photometric standard stars observed on each of the
nights of the corresponding science observations.

\section{Results}    \label{sec:results}

\begin{figure*} 
        \centering
        \epsfig{file=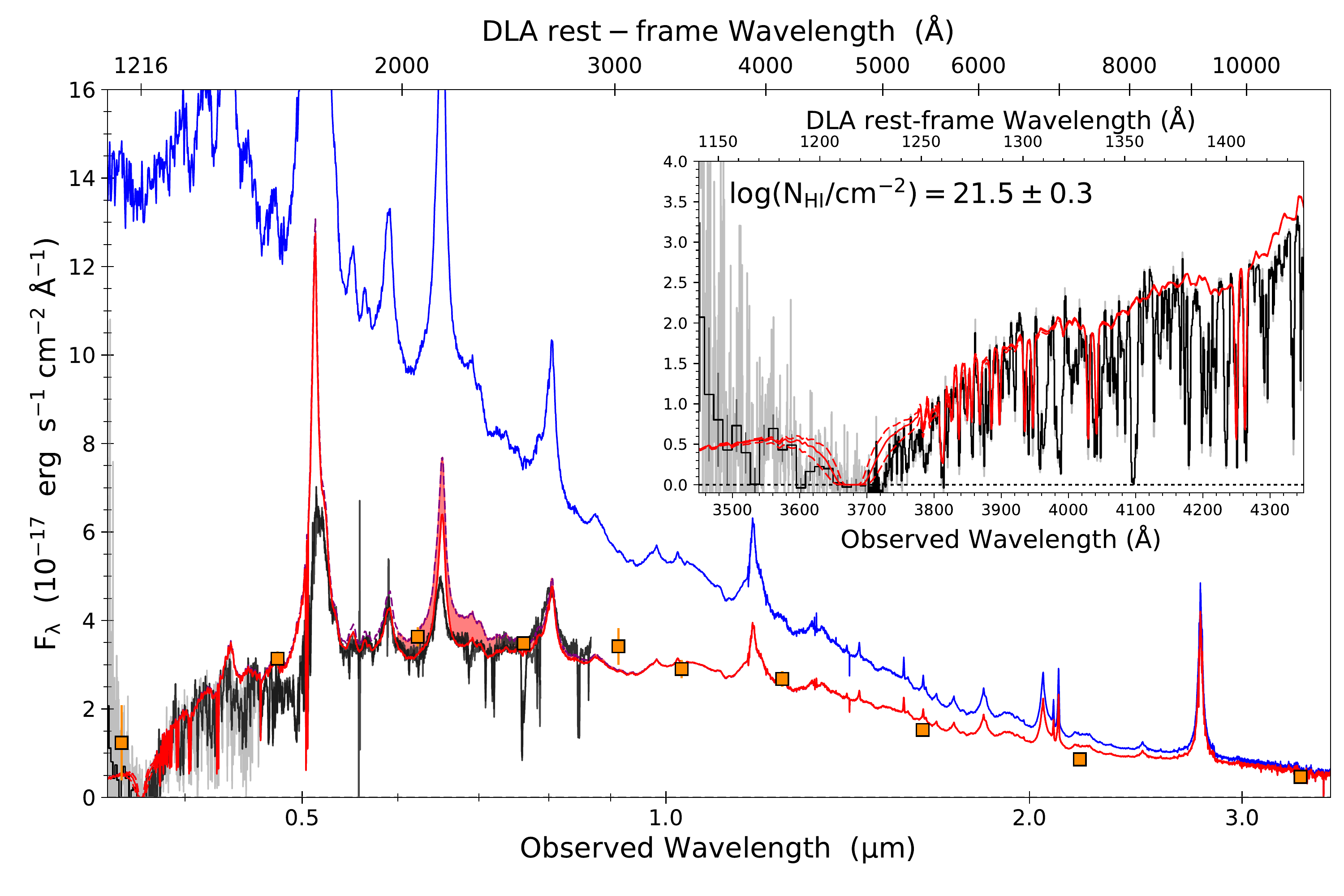,width=18cm}
    \caption{GTC spectra are shown together with the photometry from SDSS,
UKIDSS, and WISE in the $u, g, r, i, z, Y, J H, Ks$, and $W_1$ bands
\citep{Warren2007, Wright2010, Eisenstein2011}.  We also overplot a composite
quasar spectrum and a reddened composite as described in the text. The shaded
area shows the effect of the 2175-\AA \ extinction bump from dust in the
$z=2.027$ DLA galaxy. In the inset we plot a zoom on the blue part of the
spectrum where we show the 2500U spectrum covering the region around the
hydrogen Ly$\alpha$ line of the $z=2.027$ absorber. Overplotted is the DLA fit
assuming a column density of 10$^{21.5\pm0.3}$ cm$^{-2}$ ; the $\pm$ 1$\sigma$
curves are shown with dashed lines.
}
\label{fig:spectrum}
\end{figure*}

In Fig.~\ref{fig:spectrum} we show the GTC spectra together with the photometry
from SDSS, UKIDSS, and WISE. We also plot the composite quasar spectrum from
\citet{Selsing2016} and a reddened version of this quasar composite.
The overall shape of the spectrum is consistent with that found by
\citet{Krogager2016b}, i.e. with substantial reddening compared to the
unreddened quasar composite. 

\citet{Krogager2016b} found a $z=2.027$ strong metal-line absorber in the lower
resolution spectroscopy obtained at the Nordic Optical Telescope.  Based on
metal lines from \ion{Zn}{ii}, \ion{Cr}{ii}, and \ion{Fe}{ii,} in our 2500R
spectrum we determine a more precise redshift of $z_{abs} = 2.0273 \pm 0.0002$
for this system.  At this redshift lines redward of about 1700\AA \ fall on the
red side of the Ly$\alpha$ line of the quasar and hence outside of the
Lyman forest.

\subsection{Neutral hydrogen column density}

\begin{figure} 
        \centering
        \epsfig{file=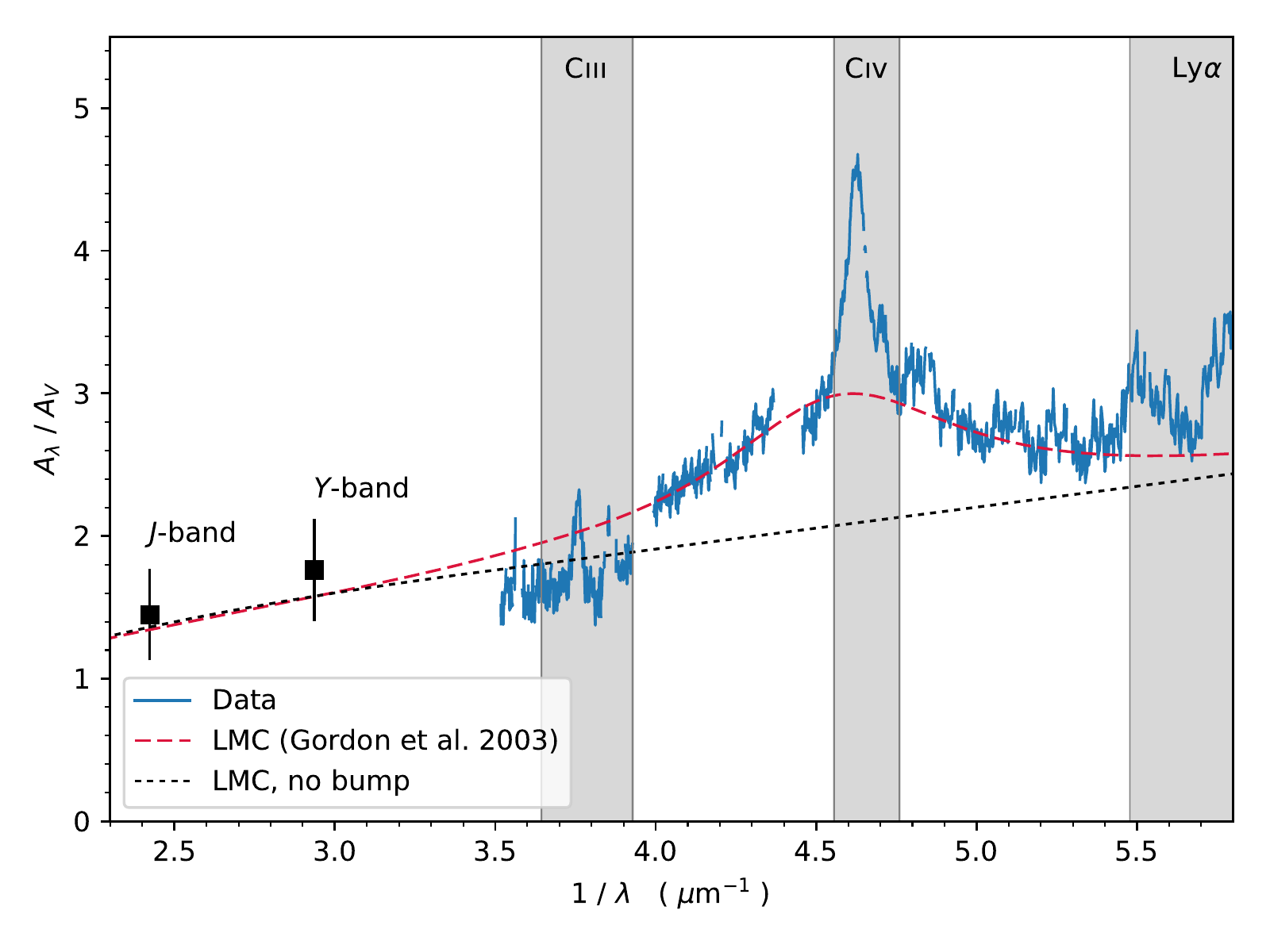,width=8cm}
    \caption{Derived extinction curve from the dust in the absorber.
    The contribution from the quasar dust has been removed to
    highlight the 2175\AA\ dust bump from the absorber. The dashed line
    shows the LMC extinction curve by \citet{Gordon2003} and the dotted
    line shows the same extinction curve but with no bump.
    The position of emission lines is indicated by the shaded bands.
    Narrow peaks are caused by differences between the intrinsic quasar
    spectrum and the template used in the analysis.
}
\label{fig:extinction}
\end{figure}

The deep 2500U spectra covers the spectral region down to and blueward of the
Ly$\alpha$ absorption line of the $z=2.027$ system. This region is shown in the
inset in the upper right-hand corner of Fig.~\ref{fig:spectrum}.  The
\ion{H}{i} column density is difficult to constrain precisely as the spectrum
of the quasar is very reddened and as the line is on the blue side of the Lyman
limit of the quasar. In the figure we have overplotted in red a model with a
quasar composite spectrum that is reddened both by dust at the quasar redshift
(SMC-like extinction at $z_{quasar}=3.214$, A$_\mathrm{V}=0.17\pm0.01$) and dust at
the DLA redshift (LMC-like extinction at $z_{DLA}=2.027$, A$_\mathrm{V}=0.22\pm0.01$).
The reddening was obtained by fitting the quasar template by \citet{Selsing2016}
to the data using a combined dust model where dust is allowed to be both at the redshift
of the absorber and the quasar. The fit is performed assuming SMC-type
dust in the quasar and LMC-type dust in the absorber owing to the presence of the
2175~\AA\ bump (for details on this dual redshift dust fitting method see 
\citet{Krogager2015, Krogager2016b}). The extinction curves
used for the analysis are taken from the parametrization by \citet{Gordon2003}.
The statistical uncertainties from the fit on the derived ${\rm A_V}$ values is 0.01~mag, however, the
uncertainty is dominated by systematic uncertainties because the intrinsic
spectral shape of the template is not known a priori. This uncertainty is approximately
0.07~mag \citep{Krogager2016a}. The extinction curve for the absorber is shown in
Fig.~\ref{fig:extinction}, where we isolated the dust contribution from the DLA.
The quasar continuum model is very similar to the best fit derived from a formal fit to independent
data from the Nordic Optical Telescope by \citet{Krogager2016b}. We
also included the effect of partial Lyman-limit absorption from two
relatively strong Lyman-forest systems at $z=3.1428$ and $z=3.1555,$ assuming
column densities of 10$^{16.4}$ cm$^{-2}$ and 10$^{16.6}$ cm$^{-2}$,
respectively, to better model the continnum around the DLA line of the $z=2.027$
absorber. Also included in the model are the Lyman lines from the two
Lyman-limit systems (up to Lyman-31) and metal lines from
\ion{S}{ii}$\lambda$1250,1253,1259, \ion{Si}{ii}$\lambda$1260, and
\ion{Fe}{ii}$\lambda$1260.  For the $z=2.027$ absorber we plotted curves
for hydrogen column densities of 10$^{21.5\pm0.3}$ cm$^{-2}$ using the
approximation of \citet{TepperGarcia2006, TepperGarcia2007}. This model
provides a reasonable fit to both the overall shape of the spectrum, photometry, and details of the region around the hydrogen line.

\subsection{Metallicity}

The metal lines associated with the $z=2.027$ absorber are very strong. 
In Table~\ref{tab:mag} we list the equivalent widths (EW) and column densities
of all the detected absorption lines at the redshift of the DLA.
As an example the rest-frame EW of the two components of the
\ion{Mg}{ii} doublet are 2.8 and 2.6 \AA, respectively. 
The redshift density of \ion{Mg}{ii} absorbers this strong or stronger 
is about 0.014 per co-moving path length based on the measurements in
\citet{Seyffert2013}. Compared to the DLAs in the very large sample 
of \citet{Noterdaeme2012}, this DLA is among the 20\% strongest \ion{Mg}{ii}
absorbers.

The resolution of the 2500R grism with the 0.8 arcsec slit we used is R=1850
around 6560 \AA. We confirm this from measurements of the widths of sky lines.
The seeing during the observations with the 2500R grism was above 1 arcsec and
hence the resolution was set by the slit width. This resolution corresponds to
160 km s$^{-1}$ in velocity space and this is too low to allow a
proper Voigt-profile fit to the absorption lines. Instead we used the
various \ion{Fe}{ii} lines available to constrain the curve of growth assuming
a single component. This method can be assumed to provide at least a robust
lower limit to the metallicity of the system \citep{Prochaska06}. This yields a
best-fit $b$ parameter of $43\pm3$~km~s$^{-1}$. Assuming the same $b$ parameter
for all low-ion species, we can determine the abundances for other lines.
The derived column densities are given in Table~\ref{tab:metal}. The
\ion{Si}{ii}\,$\lambda$\,1808, line is surprisingly strong and we consider it
likely that it is blended with an unidentified line. We hence consider
the derived Si abundance an upper limit. The abundance of zinc is derived from the
\ion{Zn}{ii}\,$\lambda$\,2026 line, which is blended with \ion{Mg}{i}. From the
lower resolution 1000R spectrum, which covers the \ion{Mg}{i}\,$\lambda$\,2852
line, we can estimate the effect of the \ion{Mg}{i} blend on \ion{Zn}{ii}. We
constrain this to be $0.2$~dex.  This has been corrected in the column
densities stated in Table~\ref{tab:metal}.

\begin{table}[!htbp]
        \centering
        \begin{minipage}{0.49\textwidth}
                \centering
                \begin{tabular}{lc}
                        \noalign{\smallskip} \hline \hline \noalign{\smallskip}
                        Ion & \emph{EW} (rest) \\
                                &    \AA        \\
                        \hline
                        \ion{Si}{ii}\,$\lambda$\,1808  & $0.76\pm0.07$ \\
                        \ion{Zn}{ii},\ion{Mg}{i}\,$\lambda$\,2026 &  $0.52\pm0.05$ \\
                        \ion{Zn}{ii}\,$\lambda$\,2026 & $0.39\pm0.05$ \\
                        \ion{Cr}{ii}\,$\lambda$\,2056 & $0.20\pm0.04$ \\
                        \ion{Zn}{ii},\ion{Cr}{ii}\,$\lambda$\,2062 & $0.49\pm0.05$ \\
                        \ion{Cr}{ii}\,$\lambda$\,2066 & $0.18\pm0.04$ \\
                        \ion{Fe}{ii}\,$\lambda$\,2249 & $0.43\pm0.03$ \\
                        \ion{Fe}{ii}\,$\lambda$\,2260 & $0.44\pm0.03$ \\
                        \ion{Fe}{ii}\,$\lambda$\,2344 & $1.37\pm0.04$ \\
                        \ion{Fe}{ii}\,$\lambda$\,2374 & $1.17\pm0.03$ \\
                        \ion{Fe}{ii}\,$\lambda$\,2382 & $1.84\pm0.03$ \\
                        \ion{Fe}{ii}\,$\lambda$\,2586 & $1.20\pm0.07$ \\
                        \ion{Fe}{ii}\,$\lambda$\,2600 & $1.80\pm0.05$ \\
                        \ion{Mg}{ii}\,$\lambda$\,2796 & $2.72\pm0.04$ \\
                        \ion{Mg}{ii}\,$\lambda$\,2802 & $2.61\pm0.04$ \\
                        \ion{Mg}{i}\,$\lambda$\,2852  & $1.24\pm0.06$ \\
                        \noalign{\smallskip} \hline \noalign{\smallskip}
                \end{tabular}
                \centering
        \caption{Subset of the absorption lines from the $z=2.027$ DLA. The 
                 \ion{Zn}{ii}\,$\lambda$\,2026 EW was corrected for the contribution
                 from \ion{Mg}{i}\,$\lambda$\,2026}
                \label{tab:mag}
        \end{minipage}
\end{table}

\begin{table}[!htbp]
        \centering
        \begin{minipage}{0.49\textwidth}
                \centering
                \begin{tabular}{lcccccccccccc}
                        \noalign{\smallskip} \hline \hline \noalign{\smallskip}
                        Element & $\log N(X)$ & [X/H]$_{\odot}$ \\
                                    & (cm$^{-2}$) &                 \\
                        \hline
                        Fe & $ 15.8\pm 0.1$ & $ -1.2\pm0.3$   \\
                        Si & $<16.7\pm 0.1$ & $<-0.3\pm0.3$  \\
                        Zn & $ 13.5\pm 0.1$ & $ -0.6\pm0.3$  \\
                        Cr & $ 13.9\pm 0.1$ & $ -1.2\pm0.3$  \\
                        Mg & $ 13.6\pm 0.1$ &                 \\
                        \noalign{\smallskip} \hline \noalign{\smallskip}
                \end{tabular}
                \centering
        \caption{Inferred column densities and metallicities from the DLA at $z=2.027$
        assuming a single component with fixed $b=43$~km~s$^{-1}$.}
                \label{tab:metal}
        \end{minipage}
\end{table}

\begin{figure} 
\centering
\epsfig{file=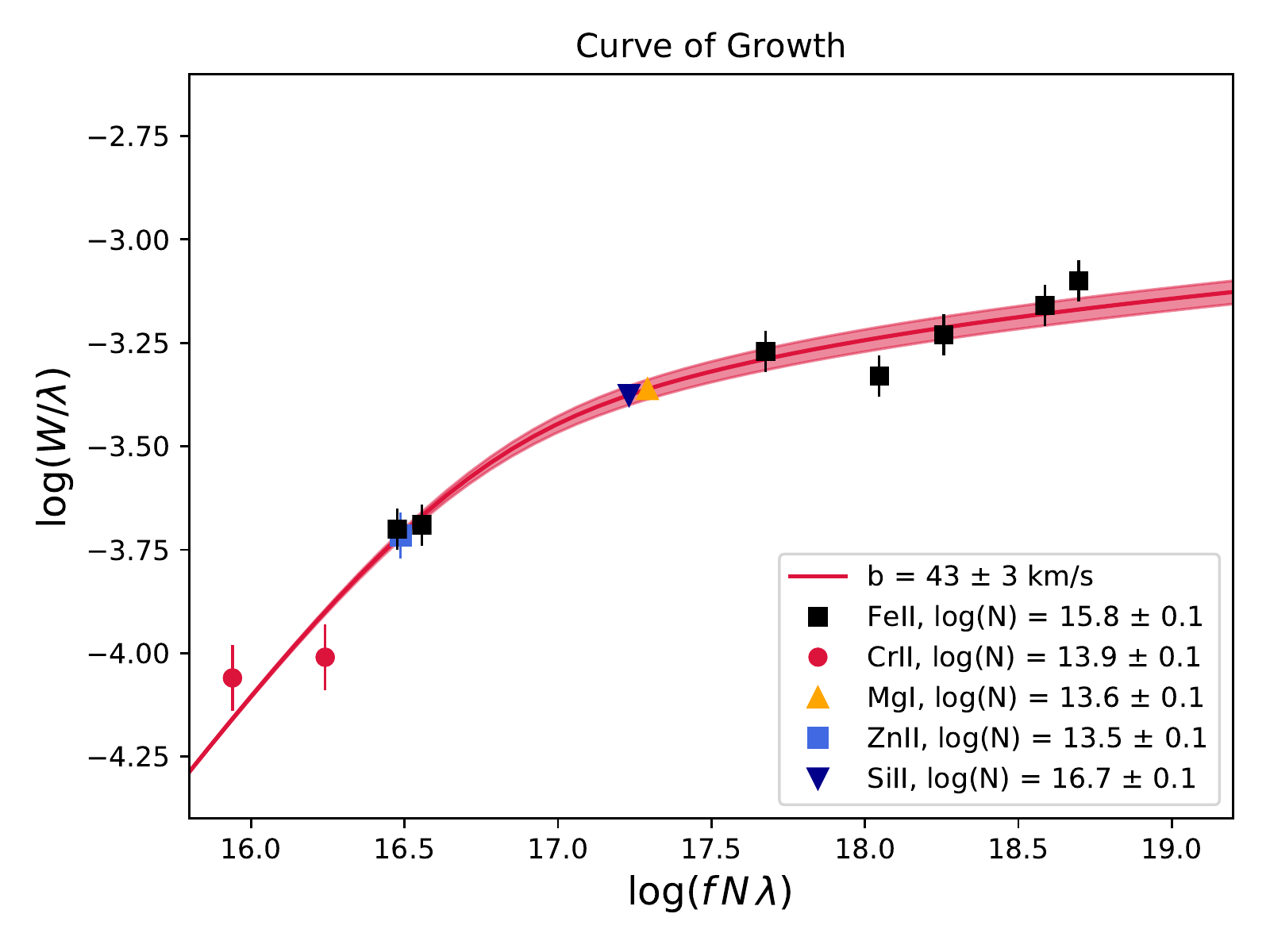,width=8cm}
\caption{Curve of growth (COG) for \ion{Fe}{ii} lines from the $z=2.027$ DLA.
The other species are assumed to follow the same COG and are plotted for comparison.
The red shaded area indicates the 1$\sigma$ uncertainty on the $b$ parameter.
}
\label{fig:metals}
\end{figure}

In Fig.~\ref{fig:metals}, we show the curve of growth (COG) derived for
the \ion{Fe}{ii} lines. Assuming a constant $b$ parameter for all other species,
we show the other available species on the same COG. For species with only one
transition available, the abundance is found by matching the equivalent width
to the COG; that is why those points are observed to lie right on the COG.

Based on the Zn measurement we can obtain the best estimate on the overall
metallicity of the system because most of the Zn should be in the gas phase \citep{Pettini1997a,
Pettini1997b}. Using solar abundances from \citet{Asplund2009}, we infer a Zn
metallicity lower limit of $-0.6\pm0.3$. The [Cr/Zn] and [Fe/Zn] ratios taken
at face value are both $-0.6\pm0.1$, which following the
relations between metallicity and depletion in \citet{DeCia2016} corresponds to
a metallicity of $-1.0$. This is at least consistent in case 
the derived column densities are close to the true values.

\subsection{Extinction}

It is not possible to derive the precise amount of extinction and
the corresponding extinction curve unambiguously. To do this requires assumptions about the
shape of the underlying quasar spectrum. The majority of bright quasars,
however, do appear to have remarkably similar spectra. This is for example
illustrated by the fact that several independent composite spectra agree well
with each other \citep[e.g.][] {Selsing2016}.

In the case of eHAQ0111+0641 \citet{Krogager2016b} assumed that the underlying,
intrinsic spectrum of eHAQ0111+0641 is similar to the composite spectrum of
\citet{Selsing2016}. They found that the spectrum was best modelled by the
composite spectrum reddened by dust both intrinsic to the quasar host galaxy
and the $z=2.027$ intervening absorber. Krogager et al. also found evidence of the
presence of the 2175-\AA \ dust extinction feature. This somewhat elusive dust
feature is well known from the Milky Way, present in the LMC, and largely
missing in extinction curves probed in the SMC. It has been detected in some
$z>1$ gamma-ray burst and quasar sightlines
\citep[e.g.][]{Junkkarinen04,Srianand2008,Eliasdottir2009,Prochaska2009,Noterdaeme2009,Conroy2010,Jiang2011,
Kulkarni11,
Zafar2012,Ma2015,Ledoux2015}.

In Fig.~\ref{fig:spectrum} we show that the model
derived by \citet{Krogager2016b} provides a very good match to our
spectra, including the 2175-\AA \ extinction feature illustrated
with the shaded area. 

\subsection{Emission from the galaxy counterpart}

We find no evidence for emission lines from the galaxy counterpart of the
DLA.  The only line our spectroscopy allows us to detect is the Ly$\alpha$
emission line and that is not detected in our 2500U grism observation down to
a 3$\sigma$ flux limit of $5\times10^{-17}$ erg s$^{-1}$ cm$^{-2}$.  This
corresponds to a star formation rate of 1.5 M$_{\sun}$ yr$^{-1}$ following the
calculation in \citet{Fynbo2002}. This non-detection is not surprising given
that we only have one slit position and given that Ly-$\alpha$ emission can
often be very weak even for strongly star-forming DLA galaxy counterparts
\cite[e.g.][]{Fynbo2011,Krogager2017}.

The target  would be very interesting for a follow-up study in the near-IR that would
allow detection of the rest-frame optical emission lines, such as [\ion{O}{ii}],
[\ion{O}{iii}], and Balmer lines. 

\section{Discussion and conclusions} 
\label{sec:conc}

\begin{figure*} 
\centering 
\epsfig{file=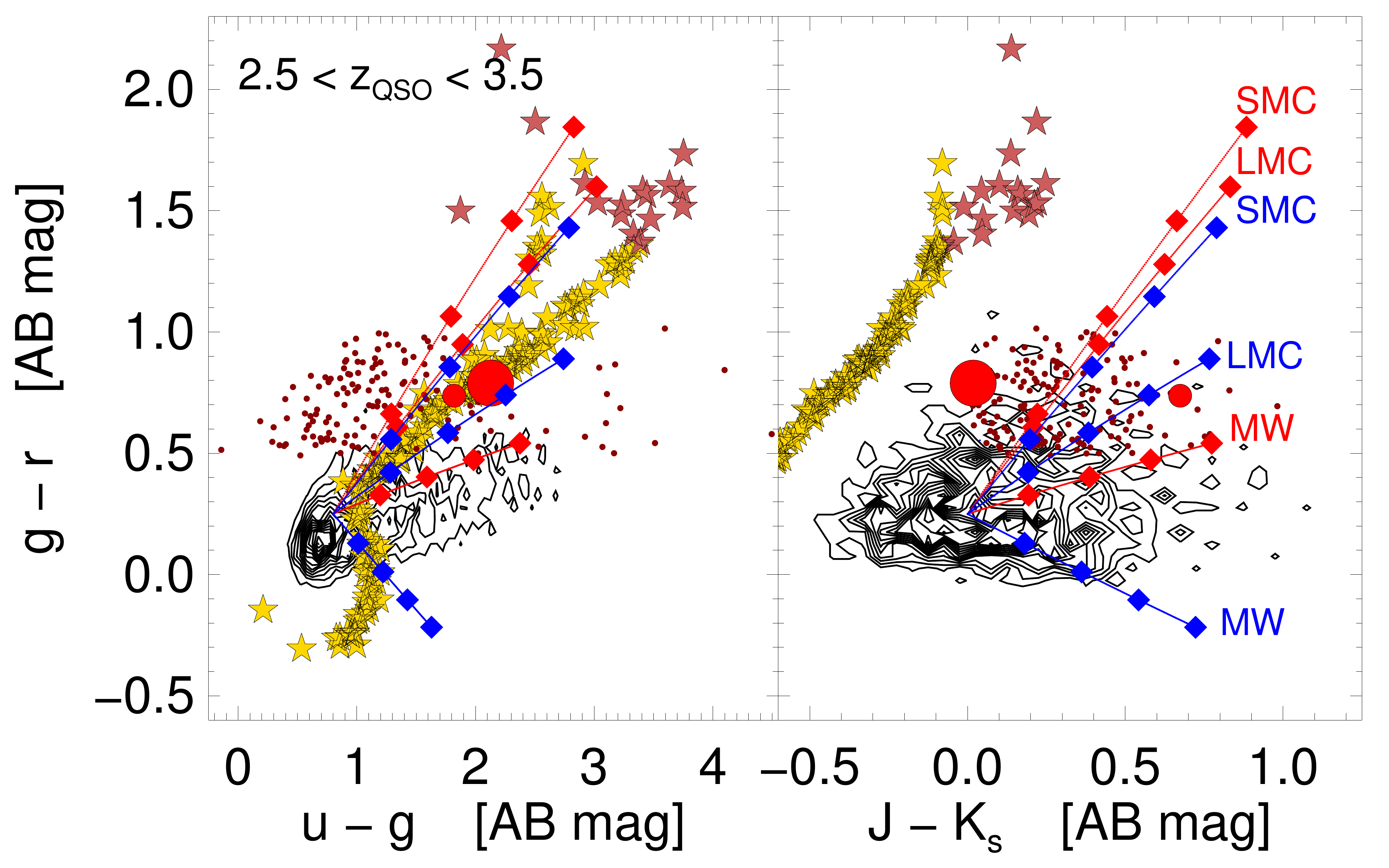,width=17cm}
\caption{Black contours shown the colour-colour distributions of the
SDSS/BOSS-DR12 quasar sample in the redshift range, $2.5 < z < 3.5$. The reddening
has moved eHAQ0111+0641 (shown with a large red circle) away from the quasar
locus onto the stellar track indicated with yellow (normal stars) and red
(M-dwarfs and later) stars. The small red circle shows the position of
HAQ2225+0527 discussed in \citet{Krogager2016b}, which also falls on the
stellar track in the SDSS colours. In the near-IR the object is well separated
from the stellar track.  The dark red points show the distribution of 150 HAQs
from \citet{Krogager2015}. Also plotted are the reddening vectors for a $z=3$
quasar reddened by SMC, LMC, and MW extinction curves from \citet{Pei1992} at
foreground redshifts of 2.0 (blue lines) and 2.5 (red lines). The diamonds
along the lines show the points corresponding to A$_\mathrm{V}$s of 0.5, 1.0,
1.5, and 2.0.
} 
\label{fig:qso} 
\end{figure*}

In Fig.~\ref{fig:qso} we show $g-r$ versus $u-g$ and $J-K_s$ colour-colour (on the
AB magnitude system) plots 
of quasars from the SDSS / Baryon Oscillation Spectroscopic Survey (BOSS) data
release (DR) 12 \citep{Eisenstein2011} and the colours of 
normal stars and cool dwarfs from \citet{Hewett2006}. The 
plot illustrates the effect of reddening on the quasar selection probability. 
In the optical the reddening has moved eHAQ0111+0641 away from the quasar locus
onto the stellar track. In the near-IR, however, the object is clearly separated
from the stellar track owing to the redder near-IR emission given its $g-r$
colour.

Three points are remarkable. First, the quasar is relatively bright (with an AB
magnitude of 18.55$\pm$0.01 in the UKIDSS K$_s$ band), which illustrates that
also bright quasars can evade selection. This point was even more clearly
illustrated in the case of HAQ2225+0527 \citep{Krogager2016a}, where the quasar
was very bright at K$_s$ = 16.15$\pm$0.01 and still evaded selection in the
SDSS/BOSS surveys.  This quasar, however, was previously detected as a bright
radio source.  Also the quasar studied by \citet{Wang2012} is intrinsically
very bright.  Second, the amount of reddening that is sufficient to remove these
objects from the selection windows is very small -- just a few tenths of a
magnitude in the DLA rest-frame V-band. And third, the metallicity of the
absorber, although it is just a lower limit, is not very large at around 25\%
solar. In the case of HAQ2225+0527 the metallicity was higher, around solar,
but the \ion{H}{i} column density was in this case smaller such that the amount
of absorption was similar, i.e. a few tenths of a magnitude {\it dust extinction} in the DLA
rest-frame V band. \citet{Boisse1998} found
that the upper envelope of DLAs in the metallicity versus \ion{H}{i} column
density plot corresponds to a few tenths of a magnitude dust extinction in the DLA
rest-frame V band. This is consistent with the findings of \citet[][see also
\citet{Barkhouse2001,Glikman2013}] {Heintz2016}, namely that only half of the
HAQ quasars in the COSMOS field were picked up by the SDSS/BOSS quasar survey
despite only modest amounts of reddening.

The few cases of DLAs causing reddening of their background quasars just
straddle this envelope noticed by \citet{Boisse1998}, but there are good
reasons to think that there are systems further above this apparent demarcation
line. From spectroscopic studies of gamma-ray burst afterglows we know of
several systems with significantly higher extinction in the rest-frame V band and
with relatively high metallicities, for example the case of GRB\,080607
\citep{Prochaska2009} or GRB\,050401 \citep{Watson2006}. \citet{Kruehler2013}
have argued that these systems probably are more common among GRB absorbers than
inferred from the samples of well-observed optical afterglow due to dust bias
\citep[see also][who find evidence for dust bias in the sample of
GRBs with detected optical afterglows]{Fynbo2009}. The reduced number of absorbers above the Boiss{\'e}-line may be
partly caused by the conversion of \ion{H}{i} to moecular hydrogen 
\citep{Schaye01,Krumholz09}.

DLAs
are a crucial class of objects in the reconstruction of the cosmic
chemical enrichment history \citep[e.g.][]{Pettini1994,Lu1996,X2003,Rafelski2014,Ledoux2015}.
\citet{Pontzen2009} find that the cosmic density of metals as 
measured from DLA surveys could be underestimated by as much as a factor of 2
owing to dust obscuration, and with the added effect of dust reddening this effect
could certainly be larger, although not dramatically larger. This is something we will try
to quantify in a future paper.
To identify more metal-rich, and hence, more dusty DLAs towards quasar sightlines,
and hence obtain a more representative sample from which to derive the cosmic
chemical enrichment history, we need to target redder (in J$-$K$_s$) and most
likely optically fainter quasars as well. In Fig.~\ref{fig:qso} we also show
colour tracks for a $z=3$ quasar spectrum reddened by extinction
curves representative for the SMC, LMC, and MW from \citet{Pei1992} with
A$_\mathrm{V}$ ranging from 0 to 2 (values at A$_\mathrm{V}$=0.5,1.,1.5, and 2.0
are indicated with diamonds) and for assumed DLA redshifts of 2.5 (red lines) and 2.0
(blue lines). It is interesting that MW-type extinction, due to the 2175\AA \
extinction feature falling in the $r$ band, makes the quasars bluer in $g-r$ 
if the DLA is at $z\approx2$. Such quasars ought to be relatively easy to identify
\citep[see also][]{Wang2004,Ledoux2015}.

In addition to reddening the extinction of course makes the quasars
fainter as well.  The upcoming EUCLID legacy survey will be a promising starting point
for such a target selection. It is already clear that the population of
reddened quasars is significant in terms of numbers.  \citet{Heintz2016} have found
from their study of a complete sample of quasars in the COSMOS field that at
magnitude limit of $J < 20$ about 20\% of quasars are reddened at a level
corresponding to E(B$-$V) > 0.1. Most likely only a few percent of these quasars are
red because of foreground DLAs, but in terms of metals this may still be an
important contribution. In a future paper we plan to make a quantitative
analysis of the magnitude and implication of the dust bias, including the
effects of both obscuration and reddening, on the measurement of the cosmic
abundance of metals in galaxies. For now we conclude that objects, such as
eHAQ0111+0641 and HAQ2225+0527, provide positive evidence that there is a dust
bias against metal-rich DLAs in existing samples of $z>2$ DLAs and that,  using 
existing samples, we
hence must to some extent be underestimating the cosmic abundances of metals in
DLAs.

\begin{acknowledgements}
Based on observations made with the Gran Telescopio Canarias (GTC), installed at
the Spanish Observatorio del Roque de los Muchachos of the Instituto de
Astrofísica de Canarias on the island of La Palma.  The research leading to
these results has received funding from the European Research Council under the
European Union's Seventh Framework Programme (FP7/2007-2013)/ERC Grant agreement
no. EGGS-278202. JK acknowledges financial support from the Danish Council for
Independent Research (EU-FP7 under the Marie-Curie grant agreement no. 600207)
with reference DFF-MOBILEX--5051-00115. KEH and PJ acknowledge support by a Project Grant
(162948--051) from The Icelandic Research Fund.  PN acknowledges support from
the {\sl Programme National de Cosmologie et Galaxies} (PNCG) funded by
CNRS/INSU-IN2P3-INP, CEA, and CNES, France.
\end{acknowledgements}

\bibliographystyle{aa}

\begin{thebibliography}{85}
\expandafter\ifx\csname natexlab\endcsname\relax\def\natexlab#1{#1}\fi

\bibitem[{{Asplund} {et~al.}(2009){Asplund}, {Grevesse}, {Sauval}, \&
  {Scott}}]{Asplund2009}
{Asplund}, M., {Grevesse}, N., {Sauval}, A.~J., \& {Scott}, P. 2009, \araa, 47,
  481

\bibitem[{{Barkhouse} \& {Hall}(2001)}]{Barkhouse2001}
{Barkhouse}, W.~A. \& {Hall}, P.~B. 2001, \aj, 121, 2843

\bibitem[{{Boiss{\'e}} {et~al.}(1998){Boiss{\'e}}, {Le Brun}, {Bergeron}, \&
  {Deharveng}}]{Boisse1998}
{Boiss{\'e}}, P., {Le Brun}, V., {Bergeron}, J., \& {Deharveng}, J.-M. 1998,
  \aap, 333, 841

\bibitem[{{Christensen} {et~al.}(2014){Christensen}, {M{\o}ller}, {Fynbo}, \&
  {Zafar}}]{Christensen2014}
{Christensen}, L., {M{\o}ller}, P., {Fynbo}, J.~P.~U., \& {Zafar}, T. 2014,
  \mnras, 445, 225

\bibitem[{{Conroy} {et~al.}(2010){Conroy}, {Schiminovich}, \&
  {Blanton}}]{Conroy2010}
{Conroy}, C., {Schiminovich}, D., \& {Blanton}, M.~R. 2010, \apj, 718, 184

\bibitem[{{De Cia} {et~al.}(2016){De Cia}, {Ledoux}, {Mattsson}, {Petitjean},
  {Srianand}, {Gavignaud}, \& {Jenkins}}]{DeCia2016}
{De Cia}, A., {Ledoux}, C., {Mattsson}, L., {et~al.} 2016, \aap, 596, A97

\bibitem[{{Djorgovski} {et~al.}(1996){Djorgovski}, {Pahre}, {Bechtold}, \&
  {Elston}}]{Djorgovski1996}
{Djorgovski}, S.~G., {Pahre}, M.~A., {Bechtold}, J., \& {Elston}, R. 1996,
  \nat, 382, 234

\bibitem[{{Eisenstein} {et~al.}(2011){Eisenstein}, {Weinberg}, {Agol},
  {Aihara}, {Allende Prieto}, {Anderson}, {Arns}, {Aubourg}, {Bailey},
  {Balbinot}, \& et~al.}]{Eisenstein2011}
{Eisenstein}, D.~J., {Weinberg}, D.~H., {Agol}, E., {et~al.} 2011, \aj, 142, 72

\bibitem[{{El{\'{\i}}asd{\'o}ttir} {et~al.}(2009){El{\'{\i}}asd{\'o}ttir},
  {Fynbo}, {Hjorth}, {Ledoux}, {Watson}, {Andersen}, {Malesani}, {Vreeswijk},
  {Prochaska}, {Sollerman}, \& {Jaunsen}}]{Eliasdottir2009}
{El{\'{\i}}asd{\'o}ttir}, {\'A}., {Fynbo}, J.~P.~U., {Hjorth}, J., {et~al.}
  2009, \apj, 697, 1725

\bibitem[{{Ellison} {et~al.}(2005){Ellison}, {Hall}, \& {Lira}}]{Ellison2005}
{Ellison}, S.~L., {Hall}, P.~B., \& {Lira}, P. 2005, \aj, 130, 1345

\bibitem[{{Ellison} {et~al.}(2001){Ellison}, {Yan}, {Hook}, {Pettini}, {Wall},
  \& {Shaver}}]{Ellison2001}
{Ellison}, S.~L., {Yan}, L., {Hook}, I.~M., {et~al.} 2001, \aap, 379, 393

\bibitem[{{Fall} \& {Pei}(1993)}]{Fall1993}
{Fall}, S.~M. \& {Pei}, Y.~C. 1993, \apj, 402, 479

\bibitem[{{Fynbo} {et~al.}(2009){Fynbo}, {Jakobsson}, {Prochaska}, {Malesani},
  {Ledoux}, {de Ugarte Postigo}, {Nardini}, {Vreeswijk}, {Wiersema}, {Hjorth},
  {Sollerman}, {Chen}, {Th{\"o}ne}, {Bj{\"o}rnsson}, {Bloom}, {Castro-Tirado},
  {Christensen}, {De Cia}, {Fruchter}, {Gorosabel}, {Graham}, {Jaunsen},
  {Jensen}, {Kann}, {Kouveliotou}, {Levan}, {Maund}, {Masetti},
  {Milvang-Jensen}, {Palazzi}, {Perley}, {Pian}, {Rol}, {Schady}, {Starling},
  {Tanvir}, {Watson}, {Xu}, {Augusteijn}, {Grundahl}, {Telting}, \&
  {Quirion}}]{Fynbo2009}
{Fynbo}, J.~P.~U., {Jakobsson}, P., {Prochaska}, J.~X., {et~al.} 2009, \apjs,
  185, 526

\bibitem[{{Fynbo} {et~al.}(2013){Fynbo}, {Krogager}, {Venemans}, {Noterdaeme},
  {Vestergaard}, {M{\o}ller}, {Ledoux}, \& {Geier}}]{Fynbo2013}
{Fynbo}, J.~P.~U., {Krogager}, J.-K., {Venemans}, B., {et~al.} 2013, \apjs,
  204, 6

\bibitem[{{Fynbo} {et~al.}(2010){Fynbo}, {Laursen}, {Ledoux}, {M{\o}ller},
  {Durgapal}, {Goldoni}, {Gullberg}, {Kaper}, {Maund}, {Noterdaeme},
  {{\"O}stlin}, {Strandet}, {Toft}, {Vreeswijk}, \& {Zafar}}]{Fynbo2010}
{Fynbo}, J.~P.~U., {Laursen}, P., {Ledoux}, C., {et~al.} 2010, \mnras, 408,
  2128

\bibitem[{{Fynbo} {et~al.}(2011){Fynbo}, {Ledoux}, {Noterdaeme}, {Christensen},
  {M{\o}ller}, {Durgapal}, {Goldoni}, {Kaper}, {Krogager}, {Laursen}, {Maund},
  {Milvang-Jensen}, {Okoshi}, {Rasmussen}, {Thorsen}, {Toft}, \&
  {Zafar}}]{Fynbo2011}
{Fynbo}, J.~P.~U., {Ledoux}, C., {Noterdaeme}, P., {et~al.} 2011, \mnras, 413,
  2481

\bibitem[{{Fynbo} {et~al.}(2002){Fynbo}, {M{\o}ller}, {Thomsen}, {Hjorth},
  {Gorosabel}, {Andersen}, {Egholm}, {Holland}, {Jensen}, {Pedersen}, \&
  {Weidinger}}]{Fynbo2002}
{Fynbo}, J.~P.~U., {M{\o}ller}, P., {Thomsen}, B., {et~al.} 2002, \aap, 388,
  425

\bibitem[{{Fynbo} {et~al.}(2008){Fynbo}, {Prochaska}, {Sommer-Larsen},
  {Dessauges-Zavadsky}, \& {M{\o}ller}}]{Fynbo2008}
{Fynbo}, J.~P.~U., {Prochaska}, J.~X., {Sommer-Larsen}, J.,
  {Dessauges-Zavadsky}, M., \& {M{\o}ller}, P. 2008, \apj, 683, 321

\bibitem[{{Fynbo} {et~al.}(1999){Fynbo}, {M{\o}ller}, \& {Warren}}]{Fynbo1999}
{Fynbo}, J.~U., {M{\o}ller}, P., \& {Warren}, S.~J. 1999, \mnras, 305, 849

\bibitem[{{Glikman} {et~al.}(2013){Glikman}, {Urrutia}, {Lacy}, {Djorgovski},
  {Urry}, {Croom}, {Schneider}, {Mahabal}, {Graham}, \& {Ge}}]{Glikman2013}
{Glikman}, E., {Urrutia}, T., {Lacy}, M., {et~al.} 2013, \apj, 778, 127

\bibitem[{{Gordon} {et~al.}(2003){Gordon}, {Clayton}, {Misselt}, {Landolt}, \&
  {Wolff}}]{Gordon2003}
{Gordon}, K.~D., {Clayton}, G.~C., {Misselt}, K.~A., {Landolt}, A.~U., \&
  {Wolff}, M.~J. 2003, \apj, 594, 279

\bibitem[{{Haehnelt} {et~al.}(2000){Haehnelt}, {Steinmetz}, \&
  {Rauch}}]{Haehnelt2000}
{Haehnelt}, M.~G., {Steinmetz}, M., \& {Rauch}, M. 2000, \apj, 534, 594

\bibitem[{{Heintz} {et~al.}(2016){Heintz}, {Fynbo}, {M{\o}ller},
  {Milvang-Jensen}, {Zabl}, {Maddox}, {Krogager}, {Geier}, {Vestergaard},
  {Noterdaeme}, \& {Ledoux}}]{Heintz2016}
{Heintz}, K.~E., {Fynbo}, J.~P.~U., {M{\o}ller}, P., {et~al.} 2016, \aap, 595,
  A13

\bibitem[{{Hewett} {et~al.}(2006){Hewett}, {Warren}, {Leggett}, \&
  {Hodgkin}}]{Hewett2006}
{Hewett}, P.~C., {Warren}, S.~J., {Leggett}, S.~K., \& {Hodgkin}, S.~T. 2006,
  \mnras, 367, 454

\bibitem[{{Jiang} {et~al.}(2011){Jiang}, {Ge}, {Zhou}, {Wang}, \&
  {Wang}}]{Jiang2011}
{Jiang}, P., {Ge}, J., {Zhou}, H., {Wang}, J., \& {Wang}, T. 2011, \apj, 732,
  110

\bibitem[{{Jorgenson} {et~al.}(2006){Jorgenson}, {Wolfe}, {Prochaska}, {Lu},
  {Howk}, {Cooke}, {Gawiser}, \& {Gelino}}]{Jorgenson2006}
{Jorgenson}, R.~A., {Wolfe}, A.~M., {Prochaska}, J.~X., {et~al.} 2006, \apj,
  646, 730

\bibitem[{{Junkkarinen} {et~al.}(2004){Junkkarinen}, {Cohen}, {Beaver},
  {Burbidge}, {Lyons}, \& {Madejski}}]{Junkkarinen04}
{Junkkarinen}, V.~T., {Cohen}, R.~D., {Beaver}, E.~A., {et~al.} 2004, \apj,
  614, 658

\bibitem[{{Krawczyk} {et~al.}(2015){Krawczyk}, {Richards}, {Gallagher},
  {Leighly}, {Hewett}, {Ross}, \& {Hall}}]{Krawczyk2015}
{Krawczyk}, C.~M., {Richards}, G.~T., {Gallagher}, S.~C., {et~al.} 2015, \aj,
  149, 203

\bibitem[{{Krogager} {et~al.}(2016{\natexlab{a}}){Krogager}, {Fynbo}, {Heintz},
  {Geier}, {Ledoux}, {M{\o}ller}, {Noterdaeme}, {Venemans}, \&
  {Vestergaard}}]{Krogager2016b}
{Krogager}, J.-K., {Fynbo}, J.~P.~U., {Heintz}, K.~E., {et~al.}
  2016{\natexlab{a}}, \apj, 832, 49

\bibitem[{{Krogager} {et~al.}(2012){Krogager}, {Fynbo}, {M{\o}ller}, {Ledoux},
  {Noterdaeme}, {Christensen}, {Milvang-Jensen}, \& {Sparre}}]{Krogager2012}
{Krogager}, J.-K., {Fynbo}, J.~P.~U., {M{\o}ller}, P., {et~al.} 2012, \mnras,
  424, L1

\bibitem[{{Krogager} {et~al.}(2016{\natexlab{b}}){Krogager}, {Fynbo},
  {Noterdaeme}, {Zafar}, {M{\o}ller}, {Ledoux}, {Kr{\"u}hler}, \&
  {Stockton}}]{Krogager2016a}
{Krogager}, J.-K., {Fynbo}, J.~P.~U., {Noterdaeme}, P., {et~al.}
  2016{\natexlab{b}}, \mnras, 455, 2698

\bibitem[{{Krogager} {et~al.}(2015){Krogager}, {Geier}, {Fynbo}, {Venemans},
  {Ledoux}, {M{\o}ller}, {Noterdaeme}, {Vestergaard}, {Kangas}, {Pursimo},
  {Saturni}, \& {Smirnova}}]{Krogager2015}
{Krogager}, J.-K., {Geier}, S., {Fynbo}, J.~P.~U., {et~al.} 2015, \apjs, 217, 5

\bibitem[{{Krogager} {et~al.}(2017){Krogager}, {M{\o}ller}, {Fynbo}, \&
  {Noterdaeme}}]{Krogager2017}
{Krogager}, J.-K., {M{\o}ller}, P., {Fynbo}, J.~P.~U., \& {Noterdaeme}, P.
  2017, ArXiv e-prints

\bibitem[{{Kr{\"u}hler} {et~al.}(2013){Kr{\"u}hler}, {Ledoux}, {Fynbo},
  {Vreeswijk}, {Schmidl}, {Malesani}, {Christensen}, {De Cia}, {Hjorth},
  {Jakobsson}, {Kann}, {Kaper}, {Vergani}, {Afonso}, {Covino}, {de Ugarte
  Postigo}, {D'Elia}, {Filgas}, {Goldoni}, {Greiner}, {Hartoog},
  {Milvang-Jensen}, {Nardini}, {Piranomonte}, {Rossi},
  {S{\'a}nchez-Ram{\'{\i}}rez}, {Schady}, {Schulze}, {Sudilovsky}, {Tanvir},
  {Tagliaferri}, {Watson}, {Wiersema}, {Wijers}, \& {Xu}}]{Kruehler2013}
{Kr{\"u}hler}, T., {Ledoux}, C., {Fynbo}, J.~P.~U., {et~al.} 2013, \aap, 557,
  A18

\bibitem[{{Krumholz} {et~al.}(2009){Krumholz}, {Ellison}, {Prochaska}, \&
  {Tumlinson}}]{Krumholz09}
{Krumholz}, M.~R., {Ellison}, S.~L., {Prochaska}, J.~X., \& {Tumlinson}, J.
  2009, \apjl, 701, L12

\bibitem[{{Kulkarni} {et~al.}(2011){Kulkarni}, {Torres-Garcia}, {Som}, {York},
  {Welty}, \& {Vladilo}}]{Kulkarni11}
{Kulkarni}, V.~P., {Torres-Garcia}, L.~M., {Som}, D., {et~al.} 2011, \apj, 726,
  14

\bibitem[{{Ledoux} {et~al.}(2015){Ledoux}, {Noterdaeme}, {Petitjean}, \&
  {Srianand}}]{Ledoux2015}
{Ledoux}, C., {Noterdaeme}, P., {Petitjean}, P., \& {Srianand}, R. 2015, \aap,
  580, A8

\bibitem[{{Ledoux} {et~al.}(2006){Ledoux}, {Petitjean}, {Fynbo}, {M{\o}ller},
  \& {Srianand}}]{ledoux2006}
{Ledoux}, C., {Petitjean}, P., {Fynbo}, J.~P.~U., {M{\o}ller}, P., \&
  {Srianand}, R. 2006, \aap, 457, 71

\bibitem[{{Lu} {et~al.}(1996){Lu}, {Sargent}, {Barlow}, {Churchill}, \&
  {Vogt}}]{Lu1996}
{Lu}, L., {Sargent}, W.~L.~W., {Barlow}, T.~A., {Churchill}, C.~W., \& {Vogt},
  S.~S. 1996, \apjs, 107, 475

\bibitem[{{Ma} {et~al.}(2015){Ma}, {Caucal}, {Noterdaeme}, {Ge}, {Prochaska},
  {Ji}, {Zhang}, {Rahmani}, {Jiang}, {Schneider}, {Lundgren}, \&
  {P{\^a}ris}}]{Ma2015}
{Ma}, J., {Caucal}, P., {Noterdaeme}, P., {et~al.} 2015, \mnras, 454, 1751

\bibitem[{{Madau} \& {Dickinson}(2014)}]{Madau2014}
{Madau}, P. \& {Dickinson}, M. 2014, \araa, 52, 415

\bibitem[{{M{\o}ller} {et~al.}(2004){M{\o}ller}, {Fynbo}, \&
  {Fall}}]{Moller2004}
{M{\o}ller}, P., {Fynbo}, J.~P.~U., \& {Fall}, S.~M. 2004, \aap, 422, L33

\bibitem[{{M{\o}ller} {et~al.}(2013){M{\o}ller}, {Fynbo}, {Ledoux}, \&
  {Nilsson}}]{Moller2013}
{M{\o}ller}, P., {Fynbo}, J.~P.~U., {Ledoux}, C., \& {Nilsson}, K.~K. 2013,
  \mnras, 430, 2680

\bibitem[{{M{\o}ller} \& {Warren}(1993)}]{Moller93}
{M{\o}ller}, P. \& {Warren}, S.~J. 1993, \aap, 270, 43

\bibitem[{{M{\o}ller} {et~al.}(2002){M{\o}ller}, {Warren}, {Fall}, {Fynbo}, \&
  {Jakobsen}}]{Moller2002}
{M{\o}ller}, P., {Warren}, S.~J., {Fall}, S.~M., {Fynbo}, J.~U., \& {Jakobsen},
  P. 2002, \apj, 574, 51

\bibitem[{{Neeleman} {et~al.}(2013){Neeleman}, {Wolfe}, {Prochaska}, \&
  {Rafelski}}]{Neeleman2013}
{Neeleman}, M., {Wolfe}, A.~M., {Prochaska}, J.~X., \& {Rafelski}, M. 2013,
  \apj, 769, 54

\bibitem[{{Noterdaeme} {et~al.}(2009){Noterdaeme}, {Ledoux}, {Srianand},
  {Petitjean}, \& {Lopez}}]{Noterdaeme2009}
{Noterdaeme}, P., {Ledoux}, C., {Srianand}, R., {Petitjean}, P., \& {Lopez}, S.
  2009, \aap, 503, 765

\bibitem[{{Noterdaeme} {et~al.}(2012){Noterdaeme}, {Petitjean}, {Carithers},
  {P{\^a}ris}, {Font-Ribera}, {Bailey}, {Aubourg}, {Bizyaev}, {Ebelke},
  {Finley}, {Ge}, {Malanushenko}, {Malanushenko}, {Miralda-Escud{\'e}},
  {Myers}, {Oravetz}, {Pan}, {Pieri}, {Ross}, {Schneider}, {Simmons}, \&
  {York}}]{Noterdaeme2012}
{Noterdaeme}, P., {Petitjean}, P., {Carithers}, W.~C., {et~al.} 2012, \aap,
  547, L1

\bibitem[{{Ostriker} \& {Heisler}(1984)}]{Ostriker1984}
{Ostriker}, J.~P. \& {Heisler}, J. 1984, \apj, 278, 1

\bibitem[{{Pan} {et~al.}(2017){Pan}, {Zhou}, {Ge}, {Jiang}, {Yang}, {Lu}, {Ji},
  {Zhang}, \& {Shi}}]{Pan2017}
{Pan}, X., {Zhou}, H., {Ge}, J., {et~al.} 2017, \apj, 835, 218

\bibitem[{{Pei}(1992)}]{Pei1992}
{Pei}, Y.~C. 1992, \apj, 395, 130

\bibitem[{{Pei} {et~al.}(1991){Pei}, {Fall}, \& {Bechtold}}]{Pei1991}
{Pei}, Y.~C., {Fall}, S.~M., \& {Bechtold}, J. 1991, \apj, 378, 6

\bibitem[{{Pei} {et~al.}(1999){Pei}, {Fall}, \& {Hauser}}]{Pei1999}
{Pei}, Y.~C., {Fall}, S.~M., \& {Hauser}, M.~G. 1999, \apj, 522, 604

\bibitem[{{Pettini} {et~al.}(1997{\natexlab{a}}){Pettini}, {King}, {Smith}, \&
  {Hunstead}}]{Pettini1997a}
{Pettini}, M., {King}, D.~L., {Smith}, L.~J., \& {Hunstead}, R.~W.
  1997{\natexlab{a}}, \apj, 478, 536

\bibitem[{{Pettini} {et~al.}(1994){Pettini}, {Smith}, {Hunstead}, \&
  {King}}]{Pettini1994}
{Pettini}, M., {Smith}, L.~J., {Hunstead}, R.~W., \& {King}, D.~L. 1994, \apj,
  426, 79

\bibitem[{{Pettini} {et~al.}(1997{\natexlab{b}}){Pettini}, {Smith}, {King}, \&
  {Hunstead}}]{Pettini1997b}
{Pettini}, M., {Smith}, L.~J., {King}, D.~L., \& {Hunstead}, R.~W.
  1997{\natexlab{b}}, \apj, 486, 665

\bibitem[{{Pontzen} {et~al.}(2008){Pontzen}, {Governato}, {Pettini}, {Booth},
  {Stinson}, {Wadsley}, {Brooks}, {Quinn}, \& {Haehnelt}}]{Pontzen2008}
{Pontzen}, A., {Governato}, F., {Pettini}, M., {et~al.} 2008, \mnras, 390, 1349

\bibitem[{{Pontzen} \& {Pettini}(2009)}]{Pontzen2009}
{Pontzen}, A. \& {Pettini}, M. 2009, \mnras, 393, 557

\bibitem[{{Prochaska}(2006)}]{Prochaska06}
{Prochaska}, J.~X. 2006, \apj, 650, 272

\bibitem[{{Prochaska} {et~al.}(2003){Prochaska}, {Gawiser}, {Wolfe}, {Castro},
  \& {Djorgovski}}]{X2003}
{Prochaska}, J.~X., {Gawiser}, E., {Wolfe}, A.~M., {Castro}, S., \&
  {Djorgovski}, S.~G. 2003, \apjl, 595, L9

\bibitem[{{Prochaska} {et~al.}(2009){Prochaska}, {Sheffer}, {Perley}, {Bloom},
  {Lopez}, {Dessauges-Zavadsky}, {Chen}, {Filippenko}, {Ganeshalingam}, {Li},
  {Miller}, \& {Starr}}]{Prochaska2009}
{Prochaska}, J.~X., {Sheffer}, Y., {Perley}, D.~A., {et~al.} 2009, \apjl, 691,
  L27

\bibitem[{{Rafelski} {et~al.}(2014){Rafelski}, {Neeleman}, {Fumagalli},
  {Wolfe}, \& {Prochaska}}]{Rafelski2014}
{Rafelski}, M., {Neeleman}, M., {Fumagalli}, M., {Wolfe}, A.~M., \&
  {Prochaska}, J.~X. 2014, \apjl, 782, L29

\bibitem[{{Rahmati} \& {Schaye}(2014)}]{Rahmati2014}
{Rahmati}, A. \& {Schaye}, J. 2014, \mnras, 438, 529

\bibitem[{{Schaye}(2001{\natexlab{a}})}]{Schaye01}
{Schaye}, J. 2001{\natexlab{a}}, \apjl, 562, L95

\bibitem[{{Schaye}(2001{\natexlab{b}})}]{Schaye2001}
{Schaye}, J. 2001{\natexlab{b}}, \apjl, 559, L1

\bibitem[{{Schneider} {et~al.}(2010){Schneider}, {Richards}, {Hall}, {Strauss},
  {Anderson}, {Boroson}, {Ross}, {Shen}, {Brandt}, {Fan}, {Inada}, {Jester},
  {Knapp}, {Krawczyk}, {Thakar}, {Vanden Berk}, {Voges}, {Yanny}, {York},
  {Bahcall}, {Bizyaev}, {Blanton}, {Brewington}, {Brinkmann}, {Eisenstein},
  {Frieman}, {Fukugita}, {Gray}, {Gunn}, {Hibon}, {Ivezi{\'c}}, {Kent}, {Kron},
  {Lee}, {Lupton}, {Malanushenko}, {Malanushenko}, {Oravetz}, {Pan}, {Pier},
  {Price}, {Saxe}, {Schlegel}, {Simmons}, {Snedden}, {SubbaRao}, {Szalay}, \&
  {Weinberg}}]{Schneider2010}
{Schneider}, D.~P., {Richards}, G.~T., {Hall}, P.~B., {et~al.} 2010, \aj, 139,
  2360

\bibitem[{{Selsing} {et~al.}(2016){Selsing}, {Fynbo}, {Christensen}, \&
  {Krogager}}]{Selsing2016}
{Selsing}, J., {Fynbo}, J.~P.~U., {Christensen}, L., \& {Krogager}, J.-K. 2016,
  \aap, 585, A87

\bibitem[{{Seyffert} {et~al.}(2013){Seyffert}, {Cooksey}, {Simcoe}, {O'Meara},
  {Kao}, \& {Prochaska}}]{Seyffert2013}
{Seyffert}, E.~N., {Cooksey}, K.~L., {Simcoe}, R.~A., {et~al.} 2013, \apj, 779,
  161

\bibitem[{{Smette} {et~al.}(2005){Smette}, {Wisotzki}, {Ledoux}, {Garcet},
  {Lopez}, \& {Reimers}}]{Smette2005}
{Smette}, A., {Wisotzki}, L., {Ledoux}, C., {et~al.} 2005, in IAU Colloq. 199:
  Probing Galaxies through Quasar Absorption Lines, ed. P.~{Williams}, C.-G.
  {Shu}, \& B.~{Menard}, 475--477

\bibitem[{{Srianand} {et~al.}(2008){Srianand}, {Gupta}, {Petitjean},
  {Noterdaeme}, \& {Saikia}}]{Srianand2008}
{Srianand}, R., {Gupta}, N., {Petitjean}, P., {Noterdaeme}, P., \& {Saikia},
  D.~J. 2008, \mnras, 391, L69

\bibitem[{{Stark}(2016)}]{Stark2016}
{Stark}, D.~P. 2016, \araa, 54, 761

\bibitem[{{Tepper-Garc{\'{\i}}a}(2006)}]{TepperGarcia2006}
{Tepper-Garc{\'{\i}}a}, T. 2006, \mnras, 369, 2025

\bibitem[{{Tepper-Garc{\'{\i}}a}(2007)}]{TepperGarcia2007}
{Tepper-Garc{\'{\i}}a}, T. 2007, \mnras, 382, 1375

\bibitem[{{Trenti} \& {Stiavelli}(2006)}]{Trenti2006}
{Trenti}, M. \& {Stiavelli}, M. 2006, \apj, 651, 51

\bibitem[{{Vladilo} \& {P{\'e}roux}(2005)}]{Vladilo2005}
{Vladilo}, G. \& {P{\'e}roux}, C. 2005, \aap, 444, 461

\bibitem[{{Wang} {et~al.}(2004){Wang}, {Hall}, {Ge}, {Li}, \&
  {Schneider}}]{Wang2004}
{Wang}, J., {Hall}, P.~B., {Ge}, J., {Li}, A., \& {Schneider}, D.~P. 2004,
  \apj, 609, 589

\bibitem[{{Wang} {et~al.}(2012){Wang}, {Zhou}, {Ge}, {Jiang}, {Lu},
  {Prochaska}, {Hamann}, {Wang}, {Wang}, \& {Yuan}}]{Wang2012}
{Wang}, J.-G., {Zhou}, H.-Y., {Ge}, J., {et~al.} 2012, \apj, 760, 42

\bibitem[{{Warren} {et~al.}(2007){Warren}, {Hambly}, {Dye}, {Almaini}, {Cross},
  {Edge}, {Foucaud}, {Hewett}, {Hodgkin}, {Irwin}, {Jameson}, {Lawrence},
  {Lucas}, {Adamson}, {Bandyopadhyay}, {Bryant}, {Collins}, {Davis}, {Dunlop},
  {Emerson}, {Evans}, {Gonzales-Solares}, {Hirst}, {Jarvis}, {Kendall}, {Kerr},
  {Leggett}, {Lewis}, {Mann}, {McLure}, {McMahon}, {Mortlock}, {Rawlings},
  {Read}, {Riello}, {Simpson}, {Smith}, {Sutorius}, {Targett}, \&
  {Varricatt}}]{Warren2007}
{Warren}, S.~J., {Hambly}, N.~C., {Dye}, S., {et~al.} 2007, \mnras, 375, 213

\bibitem[{{Watson} {et~al.}(2006){Watson}, {Fynbo}, {Ledoux}, {Vreeswijk},
  {Hjorth}, {Smette}, {Andersen}, {Aoki}, {Augusteijn}, {Beardmore}, {Bersier},
  {Castro Cer{\'o}n}, {D'Avanzo}, {Diaz-Fraile}, {Gorosabel}, {Hirst},
  {Jakobsson}, {Jensen}, {Kawai}, {Kosugi}, {Laursen}, {Levan}, {Masegosa},
  {N{\"a}r{\"a}nen}, {Page}, {Pedersen}, {Pozanenko}, {Reeves}, {Rumyantsev},
  {Shahbaz}, {Sharapov}, {Sollerman}, {Starling}, {Tanvir}, {Torstensson}, \&
  {Wiersema}}]{Watson2006}
{Watson}, D., {Fynbo}, J.~P.~U., {Ledoux}, C., {et~al.} 2006, \apj, 652, 1011

\bibitem[{{Weymann} {et~al.}(1981){Weymann}, {Carswell}, \&
  {Smith}}]{Weymann1981}
{Weymann}, R.~J., {Carswell}, R.~F., \& {Smith}, M.~G. 1981, \araa, 19, 41

\bibitem[{{Wolfe} {et~al.}(2005){Wolfe}, {Gawiser}, \& {Prochaska}}]{Wolfe2005}
{Wolfe}, A.~M., {Gawiser}, E., \& {Prochaska}, J.~X. 2005, \araa, 43, 861

\bibitem[{{Wolfe} {et~al.}(1986){Wolfe}, {Turnshek}, {Smith}, \&
  {Cohen}}]{Wolfe1986}
{Wolfe}, A.~M., {Turnshek}, D.~A., {Smith}, H.~E., \& {Cohen}, R.~D. 1986,
  \apjs, 61, 249

\bibitem[{{Wright} {et~al.}(2010){Wright}, {Eisenhardt}, {Mainzer}, {Ressler},
  {Cutri}, {Jarrett}, {Kirkpatrick}, {Padgett}, {McMillan}, {Skrutskie},
  {Stanford}, {Cohen}, {Walker}, {Mather}, {Leisawitz}, {Gautier}, {McLean},
  {Benford}, {Lonsdale}, {Blain}, {Mendez}, {Irace}, {Duval}, {Liu}, {Royer},
  {Heinrichsen}, {Howard}, {Shannon}, {Kendall}, {Walsh}, {Larsen}, {Cardon},
  {Schick}, {Schwalm}, {Abid}, {Fabinsky}, {Naes}, \& {Tsai}}]{Wright2010}
{Wright}, E.~L., {Eisenhardt}, P.~R.~M., {Mainzer}, A.~K., {et~al.} 2010, \aj,
  140, 1868

\bibitem[{{Zafar} {et~al.}(2015){Zafar}, {M{\o}ller}, {Watson}, {Fynbo},
  {Krogager}, {Zafar}, {Saturni}, {Geier}, \& {Venemans}}]{Zafar2015}
{Zafar}, T., {M{\o}ller}, P., {Watson}, D., {et~al.} 2015, \aap, 584, A100

\bibitem[{{Zafar} {et~al.}(2012){Zafar}, {Watson}, {El{\'{\i}}asd{\'o}ttir},
  {Fynbo}, {Kr{\"u}hler}, {Schady}, {Leloudas}, {Jakobsson}, {Th{\"o}ne},
  {Perley}, {Morgan}, {Bloom}, \& {Greiner}}]{Zafar2012}
{Zafar}, T., {Watson}, D., {El{\'{\i}}asd{\'o}ttir}, {\'A}., {et~al.} 2012,
  \apj, 753, 82

\end{thebibliography}
\newcommand{\noop}[1]{}

\object{eHAQ0111+0641}
\object{HAQ2225+0527}
\end{document}